\documentclass{elsart}
\usepackage{epsfig}
\usepackage{amsmath}

\begin{document}

\begin{frontmatter}

\title{Excitation of low-lying state by E3 transition 
in reaction with real photons\thanksref{th}}
\thanks[th]{Partially supported by the Deutsche
Forschungsgemeinschaft (SFB 634).}

\author[kh]{O.S.~Shevchenko},
\author[kh]{Yu.M.~Ranyuk\thanksref{cor}},
\author[kh]{A.M.~Dovbnya},
\author[kh]{V.I.~Noga},
\author[kh]{E.L.~Kuplennikov},
\author[kh]{A.A.~Nemashkalo},
\author[kh]{I.G.~Goncharov},
\author[kh]{V.N.~Borisenko},
\author[da,du]{V.Yu.~Ponomarev}

\address[kh]{National Science Center ``Kharkiv Institute of Physics and
 Technology'',\\
 61108, Kharkiv, Ukraine}

\address[da]{Institut f\"ur Kernphysik, Technische Universit\"at
Darmstadt, D-64289  Darmstadt, Germany}

\thanks[cor]{Corresponding author. 
{\it Email address}: 
ranyuk@kipt.kharkov.ua (Yu.M.~Ranyuk)}

\thanks[du]{Permanent address: Bogoliubov
Laboratory of Theoretical Physics,   
Joint Institute for Nuclear Research,
141980 Dubna, Russia}

\begin{abstract}

{\bf Abstract}

The yield of the isomeric state $^{117m}$Sn ($E_{iso}=314.58$\,keV)
has been measured in $(\gamma ,\gamma' )$ reaction 
by activation method with the bremsstrahlung end-point
energy from 2.1 to 3.0\,MeV. 
Only one intermediate state (IS) responsible for the isomer feeding
has been found.
The excitation energy of the IS $(2.25\pm 0.05)$\,MeV and 
photoproduction integral cross section $(0.022 \pm 0.002)$\,eV\,b
have been deduced.
Microscopic calculations within the Quasiparticle-phonon model have 
been performed to learn on the IS structure.
We conclude that the IS is excited in the present experiment by
the $E3-$transition.
\end{abstract}


\end{frontmatter}

Isomers in atomic nuclei are the levels with the total angular
momentum $J_{iso}$ significantly different from the one of the 
ground state $J_{g.s.}$.
They appear in the spectra due to the shell structure of a
mean field and their excitation energy in odd-mass nuclei does not 
exceed a few hundred keV. 
For these reasons, decay of isomers into the ground state is strongly
hindered and their lifetime vary for milliseconds to days depending 
on the spin difference $\Delta J_{iso} =|J_{g.s.}-J_{iso}|$.

Isomers are populated after decay of intermediate state(s) (IS) with 
the energy of 2-4~MeV and finite branching to the isomeric level.
These IS are excited, e.g., by bremsstrahlung radiation with the
end-point energy of 2-5~MeV.
The previous experiments~\cite{Pon90,PNC91,Car91,Hub93} 
have already shown that the number of
such IS which are linked to both the ground and isomeric states,
is very small, i.e. one-two states per MeV in spherical nuclei.
There was also a set of experiments in which the isomeric states
were populated in the ($\gamma$,n) reaction via excitation and 
cascade decay of the giant dipole resonance \cite{Gan96,Tso00}.

Up to now, analysis of the IS excited in the bremsstrahlung technique 
has been performed only for nuclei with $\Delta J_{iso} = 3, 4$ 
involving $E1-E2$ (or $E2-E1$) and $E2-E2$ sequences for the isomer 
population \cite{Pon90,Hub93,PNC95}.  
The best studied example is $^{115}$In \cite{PNC91,Car91,PNC95,Bel01,She05}
for which this type of experiments have been supplemented by 
nuclear resonance fluorescence (NRF)
$^{115}$In$(\gamma,\gamma' )^{115m}$In studies \cite{PNC95}.

In this letter we report our results on the isomer photoproduction
in $^{117}$Sn (the isotopic abundance is 7.68\,\%). 
It has the stable ground state with the spin and parity 
$J_{g.s.}^\pi=1/2^+$.
The isomeric state in this nucleus has the excitation energy
$E_{iso}= 314.58$\,keV and $J^\pi_{iso}=11/2^-$. 
Thus, the spin difference $\Delta J_{iso} = 5$ and at least the $E2-E3$ 
sequences is need to populate the isomer from the ground state. 
The $T_{1/2}$ value is very large for the isomeric state and 
equals to 13.60~days.

\begin{figure}
\begin{center}
\epsfig{figure=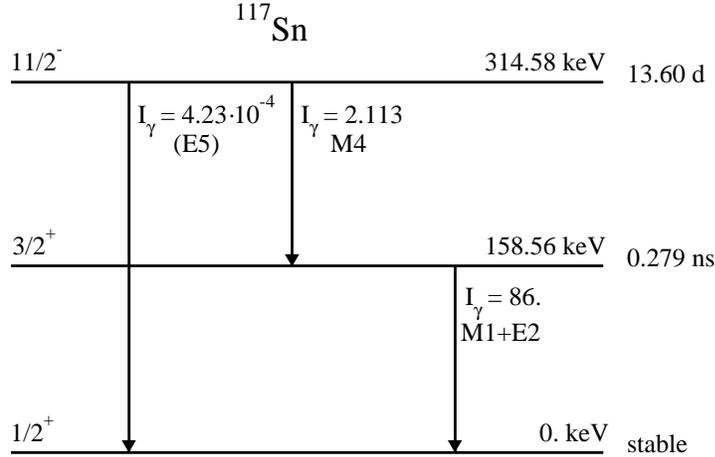,width=95mm,angle=0}   
\caption{Decay scheme of the isomeric $J^\pi_{iso}=11/2^-$ state
in $^{117m}$Sn \cite{Fir98}. See text for details.
\label{fig1}}
\end{center}
\end{figure}

The decay scheme of the isomer in $^{117}$Sn is presented in
Fig.~\ref{fig1}.
Due to the presence of the $J^\pi=3/2^+$ level at 
$E_x = 158.56$\,keV, the isomer decays predominately to this
level by the $M4$ internal conversion process with the relative 
probability ($\gamma$-line intensity) of $I_\gamma = 2.113$.
For the direct decay to the ground state, $I_\gamma = 0.000423$. 
The $158.56$~keV level decays into the ground state with  $I_\gamma = 86$
by $M1 + E2\; \gamma$-transition.

The only work in which the $^{117m}$Sn IS integrated cross
sections have been determined is Ref.~\cite{Car91}. The experiments
have been performed for the bremsstrahlung end-point energies 4 and 6 MeV. 
The obtained integrated cross sections  
$(\sigma\Gamma)$ are $(3.20\pm 0.47)$\;eV\,b and $(8.80\pm
0.26)$\;eV\,b, respectively. 
Unfortunately, low experimental resolution did not allow the IS
determination.

Our experiment have been carried out in the National Science Center
``Kharkiv Institute of Physics and Technology'' at the 3\,MV
electrostatic electron accelerator ELIAS having an energy resolution
$50$\,kV and a beam intensity up to $500\,\mu$A. 
The investigation of the isomeric states population have been performed 
using activation technique.

\begin{figure}
\begin{center}
\epsfig{figure=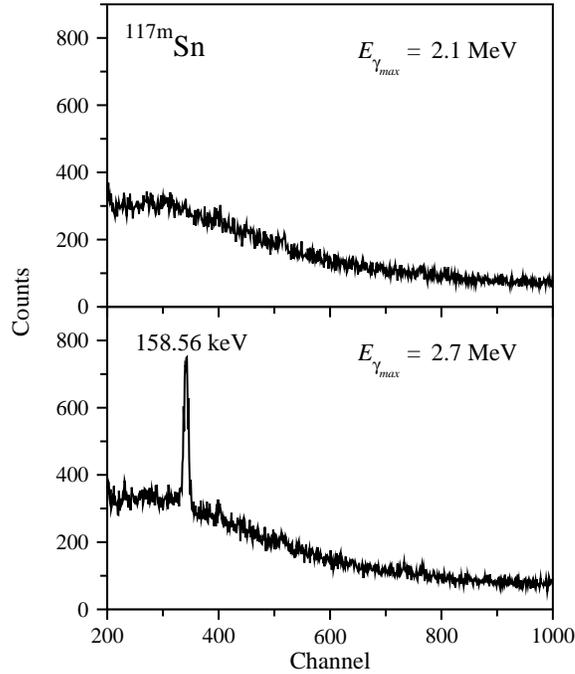,width=75mm,angle=0}   
\caption {$\gamma$-ray spectrum measured after bremsstrahlung 
irradiation of Sn target with the end-point energy of 2.1 (top)
and 2.7~MeV (bottom). 
\label{fig2}}
\end{center}
\end{figure}

Bremsstrahlung spectra have been generated by electrons irradiating of
0.5\,mm thickness Ta converter. 
Samples of natural Sn (0.2\,mm thick and 15\,mm in diameter) have been 
placed in the photon beam behind the converter. 
The photoactivities have been measured with Ge(Li) detector
by observing the $158.56$\,keV $\gamma$-rays emitted in the cascade 
of the isomeric level decay.
The detector sensitive volume is 50\,cm$^3$ and the energy resolution 
is $2.5$\,keV for 1332 keV of $^{60}$Co source.
The schematic layout of photoactivation experiment and standard
procedure for activation data development is described elsewhere
\cite{She05}. 

The obtained spectra for two different values of the end-point energy 
$E_{\gamma_{max}}$ are shown in Fig.~\ref{fig2}.
The measurements have performed with the electron beam current 
$I =  170\,\mu$A, irradiation time $t_{irr}=$ 240 and 120~min,
cooling time $t_c =$ 28 and 31~min and spectra measurement time 
$t_m = 60$\,min for $E_{\gamma_{max}} =$ 2.1 and 2.7~MeV, respectively.

The spectrum in the bottom part of Fig.~\ref{fig2} is typical for 
the end-point energy $E_{\gamma_{max}} > 2.25$\,MeV. 
The strong sharp line at 158.56~keV indicates that
the $J^\pi=3/2^+$ level at this energy is fed  after cooling 
by decay of the isomeric state. 
The absence of such a line in the top part of Fig.~\ref{fig2} means
that the isomer is not populated for $E_{\gamma_{max}} < 2.25$\,MeV.

\begin{figure}
\begin{center}
\epsfig{figure=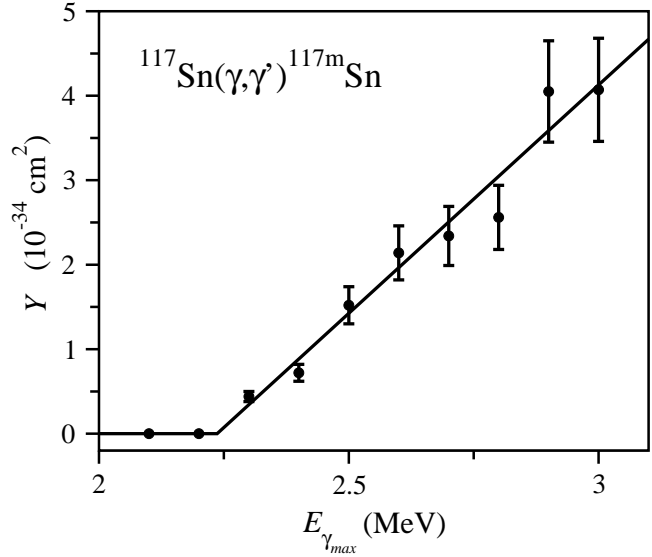,width=85mm,angle=0}   
\caption {The $^{117}$Sn isomer yield as a function of the
bremsstrahlung end-point energy. 
\label{fig3}}
\end{center}
\end{figure}

The measured isomer yield $Y(E_{\gamma_{max}})$ is presented in
Fig.~\ref{fig3} as a function of the bremsstrahlung end-point energy. 
This quantity is defined as the number of activated nuclei $N_{iso}$ 
normalized to the number of target nuclei $N_T$ per cm$^2$ and the 
number of incident electrons $N_e$:
\begin{equation}
 Y(E_{\gamma_{max}})= \frac{N_{iso}(E_{\gamma_{max}})}{N_T
 N_e}~.
\label{eq1}
\end{equation}
The error bars include systematic and statistical errors. 
Their average value is 10-15\,\%. 
The solid line in Fig.~\ref{fig3} represents the $\chi^2$ fit
of the data points assuming the linear dependence of the isomer yield
as a function of the end-point energy.

The linear dependence of the isomer yield breaks each time when 
the end-point energy of the bremsstrahlung radiation reaches the 
energy of an IS state feeding the isomer.
We conclude from the data in Fig.~\ref{fig3} that there is only
one IS of this type in $^{117}$Sn in the energy interval between
2 and 3~MeV.
The excitation energy of this state is $2.25\pm 0.05$~MeV. 

The energy of the IS in $^{117}$Sn obtained in the present experiment, 
is very close to the energy of the first 3$^-$ state in the core
$^{116}$Sn nucleus which is 2.266~MeV. 
This allows us to suppose that the IS is excited from the ground state
by the $E3$-transition. 

The isomer yield can be also calculated as 
\begin{equation}
Y(E_{\gamma_{max}})= \int_{E_{c}}^{E_{\gamma_{max}}}\sigma_\gamma
(E_\gamma ) N (E_\gamma ,E_{\gamma_{max}})d E_\gamma 
\label{eq2}
\end{equation}
where $E_c$ is a cutoff energy, $\sigma_\gamma (E_\gamma)$ is the 
reaction cross section as a function of the photon energy $E_\gamma$ 
and $N (E_\gamma ,E_{\gamma_{max}})$ represents the continuous
bremsstrahlung spectral density with the end-point energy
$E_{\gamma_{max}}$.

Accounting for the fact that only one IS has been found in
the present experiment and
assuming a small width of this level, reduces Eq.~(\ref{eq2}) to:
\begin{equation}
 Y(E_{\gamma_{max}})= (\sigma\Gamma )_{iso} \;
N (E_{IS} ,E_{\gamma_{max}})\;.
\label{eq3}
\end{equation}
The number of photons $N (E_{IS} ,E_{\gamma_{max}})$ with the 
IS energy $E_{IS}$ for each bremss\-trahlung end-point energy
in Fig.~\ref{fig3} has been calculated by mathematical modeling 
of the bremsstrahlung spectra with GEANT~3.21 program \cite{GEA}.
The number of launches was $10^7$ and we used $E_c = 0.5$~MeV and
the interval of grouping of 0.01\,MeV.

The integrated cross section for the isomer population in $^{117}$Sn
has been determined as $(0.022 \pm 0.002)$\,eV\,b.
This value is about two orders of magnitude lower than 
$(\sigma\Gamma )_{iso}$ for isomer population in nuclei with
$\Delta J_{iso} = 3$ and 4 reported in Ref.~\cite{Pon90} for $^{81}$Br,
Ref.~\cite{Hub93} for $^{89}$Y and Ref.~\cite{PNC91} for $^{115}$In.
It has been concluded from theoretical analysis that the IS in these 
nuclei are excited from the ground state either by non-collective
$E1-$ transition in $^{81}$Br \cite{Pon90} or by collective
$E2-$ transition in $^{89}$Y \cite{Hub93} and $^{115}$In \cite{PNC95}.
Very small value of $(\sigma\Gamma )_{iso}$ in $^{117}$Sn
supports our guess that the IS is excited in this nucleus by the 
$E3$-transition and calls for theoretical analysis of low-lying
states in $^{117}$Sn.
 
Such an analysis has been performed within the Quasiparticle-phonon 
model (see for details Refs.\cite{Solo92,Gale88,Brys00}) which
has been very successful in application to studies of the isomeric
state population in nuclei with $\Delta J_{iso} = 3$ and 4
\cite{Pon90,Hub93,PNC95}. 
The ground
and excited states of $^{117}$Sn have been described by the wave
function which includes quasiparticle--, [quasiparticle $\otimes$ 1
phonon]-- and [quasiparticle $\otimes$ 2 phonons]-- configurations.
The model parameters have been adjusted to reproduce the
experimental B(E2) and B(E3) values of the $2^+_1$ and $3^-_1$
states, respectively, in the neighboring even-even $^{116}$Sn core
nucleus.

The calculation reproduce very well the excitation energy of the
$3/2^+_1$ state in $^{117}$Sn (0.15\,MeV) and slightly underestimate
the energy of the isomeric $11/2^-_1$ state (0.17\,MeV). Among many
excited state in the energy interval between 1.5 and 3.5\,MeV, we
have found only one state which is linked to both, the ground and
isomeric states. It is $7/2^-$ state at the excitation energy of
2.44\,MeV with the wave function:
\begin{equation}
|7/2^-> = 0.19 \cdot 2f_{7/2} + 0.90 \cdot [3s_{1/2} \otimes
3^-_1]_{7/2^-} - 0.14 \cdot [1h_{11/2} \otimes 2^+_1]_{7/2^-} +
\ldots \label{wf}
\end{equation}
where $2^+_1$ and $3^-_1$ mean the lowest $2^+$ and $3^-$ phonons of
the core nucleus excitation.

This state is indeed excited from the ground state by the $E3$-transition 
to the second component of its wave function (\ref{wf}). Although this
is the main component of the wave function, the state decays
predominantly to the isomeric state from its third component by the
$E2$-transition with the branching ratio $\Gamma_{iso}/\Gamma_{tot}
=0.97$. The calculated value of the cross section for the isomer
population in $^{117}$Sn($\gamma, \gamma'$) reaction via this state 
is 0.026\,eV\,b and agrees very well with the experimental findings.

One finds in literature \cite{Fir98} two levels in $^{117}$Sn
at 2160 and 2280\,keV, both of them have been assigned as $5/2^-,
7/2^-$ from the $L=3$ transfer in $^{117}$Sn(p,p'). Apparently, the
above discussed $7/2^-$ state in calculation corresponds to one of
them. The calculation also yields $5/2^-$ state at 2.50\,MeV 
with the $[3s_{1/2} \otimes 3^-_1]_{5/2^-}$ configuration as the
main component in its wave function. 
But this state has no branching to the isomeric state.
Which of the levels at 2160 and 2280\,keV is our IS can be 
distinguished in the NRF experiment from the energy of the line 
at $E_{IS} - E_{iso}$.  

\begin{table}
\begin{center}
\caption{Properties of the IS in $^{117}$Sn responsible for the 
isomer feeding.}
\begin{tabular} { l c c }  \hline
&Energy (MeV) &$(\sigma\Gamma )_{iso}$ (eV b)  \\ \hline
Experiment & $2.25\pm 0.05$ &  $0.022\pm 0.002$ \\ \hline 
Theory   & $ 2.44$ &   0.026 \\ \hline
\label{tab}
\end{tabular}
\end{center}
\end{table}

To conclude, the population of the isomer in $^{117}$Sn has been
studied in reaction with the bremsstrahlung radiation with the
end-point energy from 2.1 to 3.0~MeV. 
Only one intermediate state responsible for the isomer feeding
has been found in this energy interval. 
The properties of this state from the present experiment and
theoretical analysis are summarized in Table~\ref{tab}.
In our opinion, we have convincing evidence that this IS is
excited in this extremely selective reaction by the $E3-$transition 
from the ground state.

\end{document}